\documentstyle[epsf,12pt]{article}
\begin{document}
\renewcommand{\baselinestretch}{1.5}

\newcommand\beq{\begin{equation}}
\newcommand\eeq{\end{equation}}
\newcommand\bea{\begin{eqnarray}}
\newcommand\eea{\end{eqnarray}}

\newcommand\Sum{\sum}
\newcommand\Sumi{\sum_i^N}
\newcommand\Sumj{\sum_j^D}
\newcommand\Sumij{\sum_{i,j}^N} 
\newcommand\Sumijk{\sum_{i,j\ne i,k\ne i}^N}
\newcommand\sumijk{\sum_{i\ne j\ne k}^N}
\newcommand\Pij{\prod_{i<j}^N}

\newcommand\expo{e^{-{1\over 2}\Sumi\Sumj x_i^{j2}}}
\newcommand\expon{e^{-{1\over 2}\Sumij x_i^{j2}}}
\newcommand\partialxij{\partial f/\partial x_i^j}
\newcommand\pxijt{{\partial^2 f/\partial x_i^{j2}}}
\newcommand\prl{Phys. Rev. Lett.}
\newcommand\prb{Phys. Rev. {\bf B}}

\centerline{\bf Transport in Luttinger Liquids}
\vskip 1 true cm

\centerline{Sumathi Rao \footnote{{\it e-mail
address}: sumathi@mri.ernet.in}} 
\centerline{\it Mehta Research Institute, Chhatnag Road,Jhunsi,}
\centerline{\it Allahabad 221506, India}
\vskip 2 true cm
\noindent {\bf Abstract}
\vskip 1 true cm

We give a brief introduction to  
Luttinger liquids and to the phenomena
of electronic transport or conductance in quantum wires. We explain
why the subject of transport in Luttinger liquids is relevant and
fascinating and review some  important results on tunneling through
barriers in a one-dimensional quantum wire and the phenomena of
persistent currents in mesoscopic rings.  We give a brief
description of our own work on transport through doubly-crossed
Luttinger liquids and transport in the Schulz-Shastry exactly solvable
Luttinger-like model. 

\vskip 1 true cm


\newpage

I am very happy to be here at the symposium 
held to felicitate Prof. Rajaraman, who has taught me many things in
quantum field theory as well as condensed matter physics.
\footnote{Talk presented at the  `Symposium on  Quantum Many-Body
Physics', held at Jawaharlal Nehru University, New Delhi, March 5-7, 1999.}
Since Prof. Rajaraman has had contributions in various diverse fields
of physics, such as nuclear physics, particle physics, formal field
theory and condensed matter physics, the audience here is also varied and
has representatives from all fields. Hence, I will start my talk by
first giving a brief introduction to the words such as Luttinger
liquid and transport in the title of my talk.  I will explain why the
field is both very important at the current time and at the same time
theoretically fascinating. I will then give a quick review of some of the
important results, before I go on to describe some work
that we have done and are doing in this field.

\vskip 0.3cm

\centerline{\bf I. Introduction}

\vskip 0.2cm

\noindent {\bf What is a Luttinger liquid?}

\vskip 0.3cm 

Let us first remind ourselves of a Fermi liquid. Usual condensed
matter systems deal with a collection of electrons, which are
fermions. If the fermions are non-interacting, we have a Fermi gas,
with single particle eigenstates, which can be filled upto the Fermi
level. Excitations over the ground state are quasiparticles ( above the
Fermi surface)  and quasiholes ( below the Fermi surface),
which have the same quantum numbers as that of the original 
electrons or holes.
The idea behind Fermi liquid theory, is that interactions can change
the ground state, modify the excitations  and their energies and so on, but
essentially, one continues to have single-particle fermion like
excitations even after inclusion of the interactions. 
These excitations (called Landau quasiparticles) 
can have their masses, couplings, etc
renormalised, but basically each state is in one-to-one correspondence
with the non-interacting states. Such a system is called  a Fermi
liquid system.

The Luttinger liquid\cite{LUTT} on the other hand, is the ground state of an
interacting system which no longer has quasiparticles similar to that
of the non-interacting case. Instead, it has collective excitations,
which bear no resemblance to the original fermions - they are
bosonic. Also, for a fermion, its charge and spin move together. In a
Luttinger liquid, the charge and spin degrees of freedom move
independently.  At a more technical level, instead of a pole in the
single particle propagator, even when interactions are included, as one
would expect for a Fermi liquid, here one finds
anomalous non-integer exponents. These anomalous exponents in various
correlation functions or response functions is the hallmark of
Luttinger liquid behaviour.

In three dimensions, most electronic phenomena can be understood
within the framework of Fermi liquid theory. Two dimensions is still a
doubtful case, where for some phenomena,  it is not clear
whether Fermi liquid theory is really 
applicable. For instance, many people believe that high $T_c$
superconductivity needs non-Fermi liquid behaviour. But in
one dimension, it is well-known that Fermi liquid theory breaks down
and  hence the
relevance of Luttinger liquid theory has been understood for
quite a while.

\vskip 0.3cm

\noindent {\bf Transport}

\vskip 0.3cm

Transport is an important concept in condensed matter physics. Here,
by transport, we mean electronic transport or conductance. We apply a
voltage across a wire and measure the current through it. This 
gives us the conductance. The aim is to compute the conductance as a
function of the voltage, temperature, presence of impurities or
disorder and so on. Normally, when currents are measured in wires, one
does not worry about quantum effects, because wires are still
macroscopic objects. But here, we shall be talking about
one-dimensional `mesoscopic ' wires, so quantum effects will be important. 
In fact, whenever the physical dimensions of the conductor becomes
small, (it need not be really one-dimensional), the usual Ohmic
picture of conductance where the conductance is given by
\beq
G=\sigma ~~{W\over L} =\sigma ~~{{\rm width ~~of ~~conductor}\over {\rm length
~~of ~~conductor}}
\eeq
where $\sigma$ is a material dependent quantity, breaks down. A whole
new field called `mesoscopic physics'\cite{DATTA}  
has now been created to deal 
with electronic transport in such systems. 
The term `mesoscopic' in between microscopic and
macroscopic is used for systems, where the sizes of the devices are
such that it is comparable with a) the de Broglie wavelength ( or
kinetic energy) of the electron, b) the mean free path of the electron
and c) the phase relaxation length ( the length over which the
particle loses memory of its phase) of the electron. For a macroscopic
object, the size is much larger than any of these lengths. These
lengths actually vary greatly depending on the material and also on
the temperature. Typically, at low temperatures, they vary between a
nanometer for metals to a micrometer for quantum Hall systems. 

For mesoscopic wires, in general, quantum effects need to be taken
into account. The conductances are computed using the usual
quantum mechanical formulation of transmission and reflection through
impurities. This formulation is called the Landauer-Buttiker
formulation and works for Fermi liquids. However, when we really go to
one dimensional wires, interactions change the picture dramatically,
since the quasi-particles are no longer fermion-like. Hence the
Landauer-Buttiker formalism cannot be applied and one needs to compute
conductances in Luttinger wires taking interactions into account right
from the beginning. We shall review the theoretical results of
transport in Luttinger wires after giving a brief motivation as to why
the study of transport through  Luttinger wires is interesting.

\vskip 0.3cm

\centerline{\bf Motivation}

\vskip 0.3cm

The main motivation in this field is that
recently, advances in nanotechnology, and the discovery of new
one-dimensional materials such as carbon nanotubes have enabled the
fabrication of extremely narrow wires\cite{1DWIRE}. 
Experiments\cite{YACOBY} of fundamental theoretical
importance have been performed on these wires such as
those that look for coherent scattering  and 
measure the phase of the transport of the
electron through barriers in these wires. One can also try to 
look for Luttinger
liquid behaviour in these wires by measuring 
their transport properties. However, it is hard to observe
Luttinger liquid behaviour because any residual disorder or 
any deviation from 
one dimensionality, obscures the power laws which are
characteristic of Luttinger liquids.

Some of the experiments\cite{1DWIRE} looking for 
Luttinger liquid behaviour include

\begin{itemize}

\item{} Experiments on semi-conductor wires.

These are quantum wires because they are of mesoscopic sizes and are
at low temperatures. But it not yet clear whether Luttinger liquid
behaviour has been seen in these wires or whether the experiments can
be explained by Fermi liquid theory.

\item{} Tunneling into edge states in the Fractional Quantum Hall
effect. 

Here, Luttinger liquid behaviour has actually been seen. For Fermi
liquid behaviour (which is seen for the $\nu=1$ state), $I\propto
V$. But for Luttinger liquids, the exponent changes. It was
theoretically predicted to be  $I/V \sim  V^2$ and 
$G \sim T^2$ for $\nu=1/3$ state, and
experimentally found to be $I/V \sim  V^{1.7 \pm 0.06}$ and 
$G \sim T^{1.75\pm
0.08}$, where $G$ is the conductance as a function of the temperature.

\item{} Experiments on single carbon nano-tubes.

Transport along a single carbon nano-tube has been experimentally
measured and the results have been similar to those for semi-conductor
wires. 

\end{itemize}

There have also been predictions\cite{PRED} 
that armchair nanotubes form a Luttinger
liquid and the appropriate power laws for various conductances have
been calculated. But the main point that I wish to emphasize is that
at the moment, several experiments are being done on 1-D
systems. Since,in 1-D systems, interactions exist and change the
physics drastically, it becomes important to take it into
account. Theory can hence lead to predictions which can be immediately
tested. At a deeper level, the field involves a fascinating interplay
of concepts from strong correlations, impurities and disorder as well
as mesoscopic systems. 

\vskip 0.3cm
\centerline{\bf  Review of Important Results} 
\vskip 0.3cm

I will now review a couple of important results in the field which are
required to explain my work. 

\noindent {\bf Persistent currents}

In the presence of an external magnetic field, it is possible to have
persistent currents\cite{PERSISTENT} in small metal rings.
This is a very simple quantum mechanical phenomena, which can be
understood on the basis of the Aharanov-Bohm effect. The idea is that
we have a small metal ring of circumference $L$ 
and thread a magnetic flux through it. Hence, the wave-function of an
electron that goes along the ring, picks up a phase $\psi(x+L) =
e^{2i\pi \phi/\phi_0} \psi(x)$ after it completes a circuit.
Since the sign of the phase is different for left-movers and
right-movers, this breaks the degeneracy between them on the
ring. Thus, for a given chemical potential, there are more rightmovers
than left (or vice-versa), which leads to a current. One can compute
the current by adding up the contribution of all the levels below the
Fermi level  and plot it against the flux to get the sawtooth 
picture depicted in Fig.(1). The periodicity in $\phi/\phi_0$ 
is clearly because the flux
$\phi_0$ where $\phi_0 = hc/e$ is the unit of flux is
indistinguishable from  no flux.

\epsfxsize=3.6 in
\epsfysize=2.0 in
\begin{center}
\epsffile{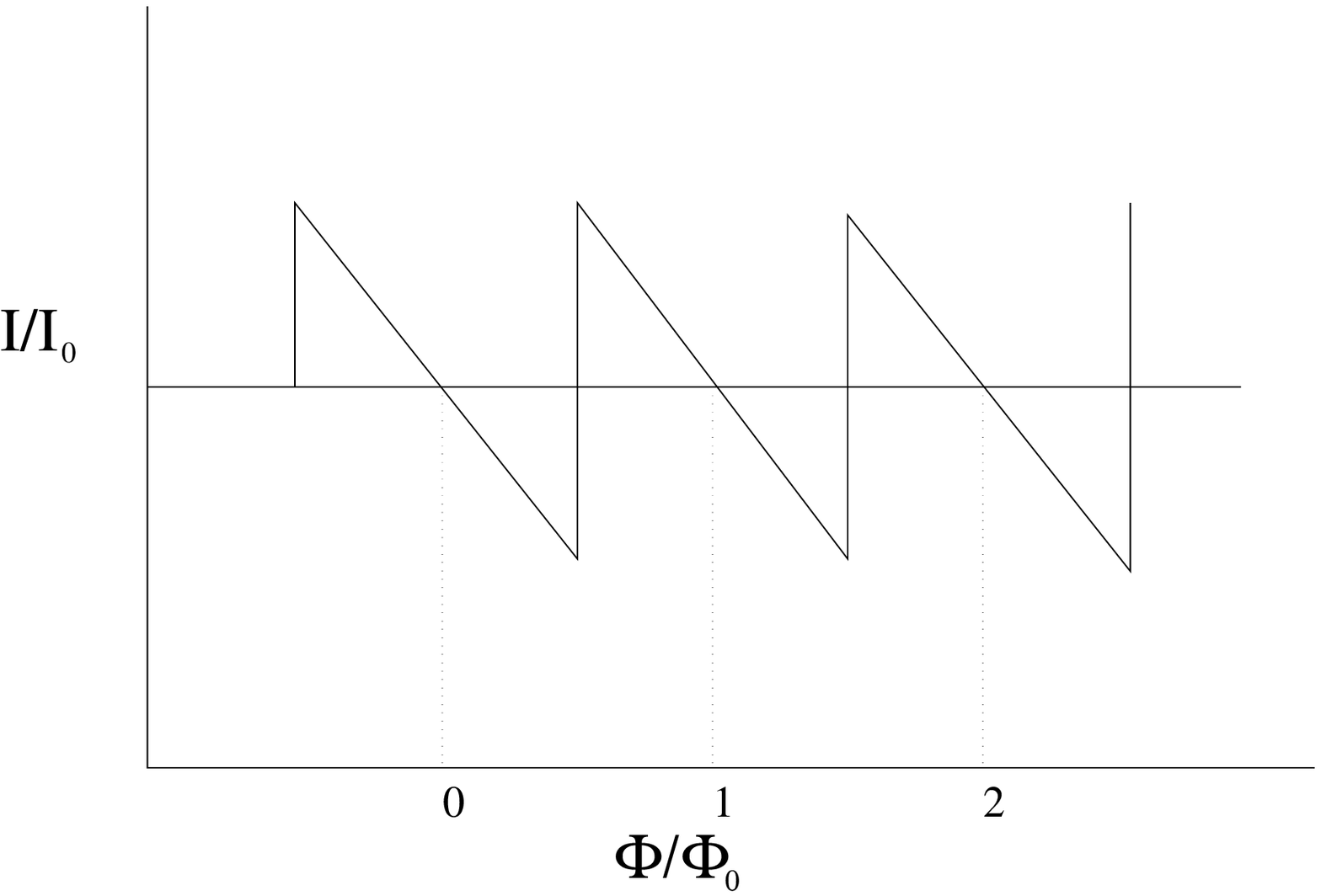}
\end{center}
\begin{itemize}
\item{ \bf Fig 1. } 
Current versus flux through the ring. The periodicity for
$\phi=n\phi_0$ is clearly visible.
\end{itemize}

\noindent {\bf Kane-Fisher results}

Kane and Fisher in a pathbreaking paper\cite{KF} in 1992, showed how
interactions in a one-dimensional system, changed transmission through
barriers dramatically. We shall describe the idea behind their work in
some detail here.

The simplest Luttinger liquid model can be described by two bosons,
one for the charge and one for the spin degree of freedom. The action
(in one space, one time dimension) is given by
\beq
S=\int dx d\tau {\large (}{v_{\rho} g_{\rho}\over 2} [(\nabla \phi_{\rho})^2 + 
{1\over v_\rho^2} (\partial_{\tau} \phi_{\rho})^2 ] +
{v_{\sigma} g_{\sigma}\over 2}[(\nabla \phi_{\sigma})^2 + 
{1\over v_\sigma^2} (\partial_{\tau} \phi_{\sigma})^2 ]{\large )}
\eeq
where the field $\phi_{\rho}$ and $\phi_{\sigma}$ denote the charge
and spin degrees of freedom respectively and the velocities of the two
fields can be different, since spin and charge degrees of freedom
decouple in a Luttinger liquid. $g_{\rho}$ and $g_{\sigma}$ are
parameters that are a measure of the strengths of interaction of the
original fermions.   
Equivalently, the $g$-parameters can be identified with 
the radii of compactification $R$ of the free
bosons of a $c=1$ conformal field theory, 
with the free fermion point being identified with a particular
value of $R$. The point to note is that these parameters do not
denote interactions in the bosonic model. $g_{\rho}=2$ ($g_{\rho}=1$ for
spinless fermions) denotes the
free fermion point for the charge degree of freedom. 
Generically, unless there is a magnetic field,
$g_{\sigma}=2$ in order to repect the $SU(2)$ symmetry. 

So now, we have a one-dimensional wire made up of Luttinger
bosons. What happens when we put a voltage across the wire and measure
the current? Initially, Kane and Fisher claimed that the conductance
depended on the Luttinger parameter and was given by $I/V = g_{\rho}
e^2/h$, but now, it is generally accepted\cite{PURE} that for a pure Luttinger
wire, $I/V = e^2/h$, just as it is for a non-interacting Fermi liquid
wire. The reason for this is that if the wire has no impurity, then
its only resistance comes from the contacts at the leads, - the
one-dimensional wire is connected to three-dimensional Fermi liquid
leads at the two ends of the wire. Since the resistance comes only
from the contacts, it does not matter whether the fermions in
the wire are interacting or not.  This explains why the conductance is the
same as that for non-interacting fermions.
 
Now, we may ask : `What happens if we introduce an impurity?'
(The impurity may be a barrier, a constriction or a localised
impurity.) The impurity is modelled by a potential $V(x)$ at or around
the origin, so that the Hamiltonian (for spinless fermions) is modified by 
\beq
\delta H = \int dx V(x) \psi^{\dagger}(x) \psi(x)
\eeq
in the weak barrier limit. In the strong barrier limit,
it is more appropriate to think of two independent wires to the left
and right of the origin  - $i.e.$, a wire which is cut at the origin -
and then allow a small hopping, given by
\beq
\delta H = -t[\psi^{\dagger}_+(x=0) \psi_-(x=0) + h.c.].
\eeq
In either case, we can bosonise these terms and use perturbation
theory in $V$ or $t$ to obtain the renormalisation group equations
given by
\beq
{dV\over dl} = (1-g) V  \quad {\rm or} \quad {dt\over dl} = (1-{1\over
g}) t.
\eeq
Hence, for $g<1$, the barrier term is relevant and grows ( and from
the other limit,
the hopping term is irrelevant and becomes weak). So
for $g<1$, either from the weak coupling or the strong coupling side,
the result is that the impurity `cuts' the wire and there is no
transmission. For $g>1$, on the other hand, $V$ decreases and $t$
increases. In other words, even if we start with a barrier, it
vanishes under RG and transmission becomes perfect - $i.e$, the wire
is `healed'.

$g<1$ corresponds to repulsive interactions in the original fermionic
model, whereas $g>1$ corresponds attractive interactions. 
$g=1$ is the free fermion limit
for spinless fermions, where both the perturbing operators, either from
the weak barrier or strong barrier side,  are marginal. Here, as we
know from usual one-dimensional quantum mechanics, one can have both
transmission and reflection. Similar results can also be found for
electrons with spin.

The next thing is to study transmission through two barriers. For
$g<1$, one may expect that since even one barrier cuts the wire, there
is no chance of transmission. However, surprisingly, it is still
possible to have resonant transmission. The idea is that one can have  
quasi-bound states between the barriers, which will correspond to the
energies for resonant transmission. For weak barriers, this happens at
the energies at which backscattering from both sides 
can be tuned to be zero. For
strong barriers, this happens when the energy on the island between
the two barriers is degenerate for two states, which is again arranged
by tuning the chemical potential  
on the island, so that the energy to add another
electron is zero. Hence, one finds resonances as a function of the
gate voltage.  

This is somewhat reminiscent of what is called Coulomb blockade physics
for non-interacting electrons. Even for non-interacting electrons, the
mesoscopic length scale of the island, implies that it has a small
capacitance - it can only hold so much charge. To add another electron
to the island costs Coulomb charging energy $e^2/C$, where $C$ is the
capacitance of the island. (Note that this is different from  the
interaction represented by the Luttinger parameter $g<1$, which is
only a measure of the short-range part of the repulsive interaction
between electrons.) If the charging energy can be neutralised by
changing the chemical potential on the island by a gate voltage, then
there is no energy required to add another electron and one can get
resonant transmission. This can only happen for specific values of
the gate voltage. At other values of the voltage, there is no transmission
because the Coulomb energy blocks the passage of the electron. Hence, one
gets peaks in the conductance at particular values of the gate voltage
or equivalently, one 
gets plateaux and jumps in the graph of the current versus the gate
voltage, which is called the Coulomb staircase. This is depicted in
Fig. (2).

\epsfxsize=4.2 in
\epsfysize=2.0 in
\epsffile{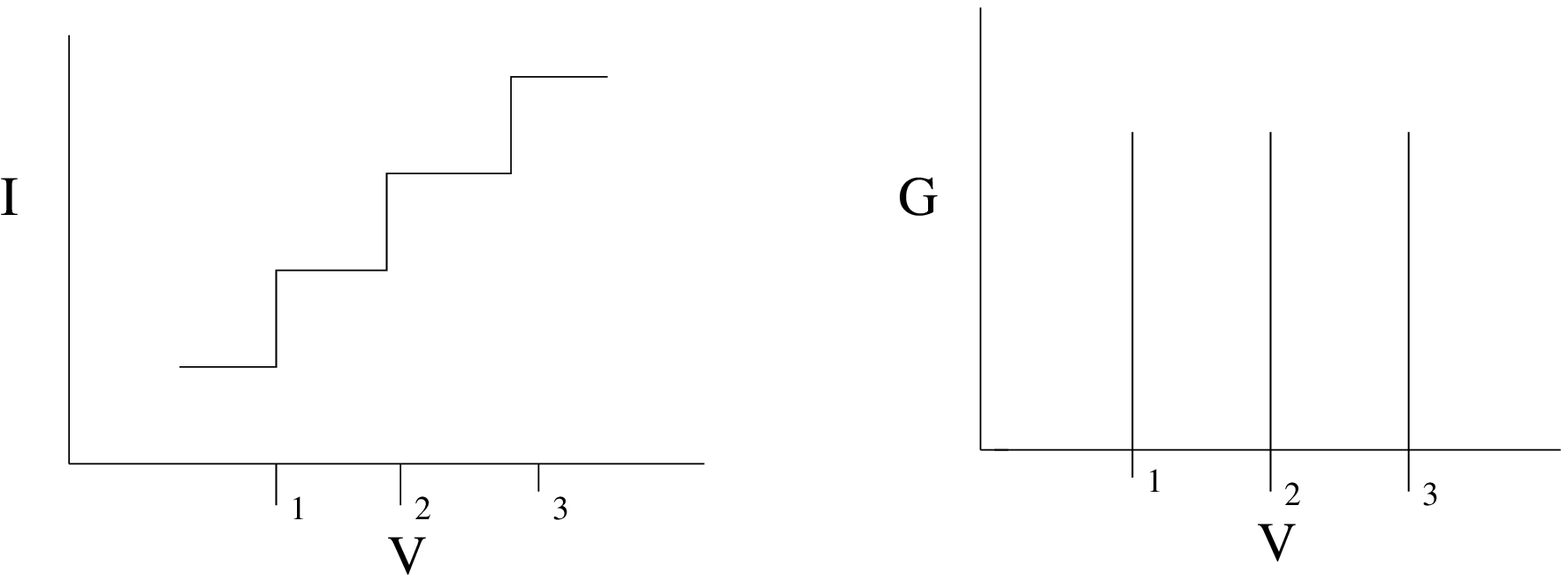}
\begin{itemize}
\item{ \bf Fig 2. } 
Current $I$ versus voltage $V$ and conductance $G$ versus $V$ at zero
temperature 
for a Luttinger liquid wire with two barriers. Note the sharpness of
the jumps, which is the feature that sets it apart from the analogous
Coulomb blockade for non-interacting electrons.
\end{itemize}

\noindent However, the physics which causes the resonances in the interacting
model is analogous but not identical to the Coulomb blockade
physics. Unlike for the non-interacting case, where there is a finite
width to the resonances, for the Luttinger liquid model, the
resonances become infinitely sharp at zero temperature.

\vskip 0.3cm

\centerline{\bf Our work} 

\vskip 0.3cm

Here, we will report briefly on two pieces of work, which uses some of
the results that we have reviewed above. One of them is on
transmission through a particular geometry of Luttinger liquid
wires, which is interesting - that of doubly crossed Luttinger
wires\cite{PDN}. The second, which is still incomplete, 
is on some exact results
in a toy model of Luttinger liquid\cite{PEEKAY}. 

\noindent {\bf Double-crossed Luttinger wires}

One motivation for studying crossed Luttinger wires 
was that in the standard two Luttinger chain
problem, with couplings all along the wire between the two chains, the
coupling was relevant and led to a flow away from the Luttinger liquid
fixed point of a single chain. So our  aim was to try and include
couplings between chains at several points and see whether Luttinger
liquid behaviour is destroyed. But interestingly, point-like couplings
even for one and two points lead to unusual transport features. 
Even for just a single crossing of two Luttinger liquids, it was
found\cite{KE} 
that the current in one wire was extremely sensitive to the voltage
drop across the other wire. Here, we study two Luttinger liquid wires 
coupled at two points and
connected to external constant voltage sources. The aim was to see
whether one can tune resonant transmission in both wires by applying
gate voltages. Our conclusion was that in the the limit when the
external biases tended to zero, a single gate voltage was sufficient
to tune for resonances in both wires.

\epsfxsize=3.2 in
\epsfysize=2.0 in
\begin{center}
\epsffile{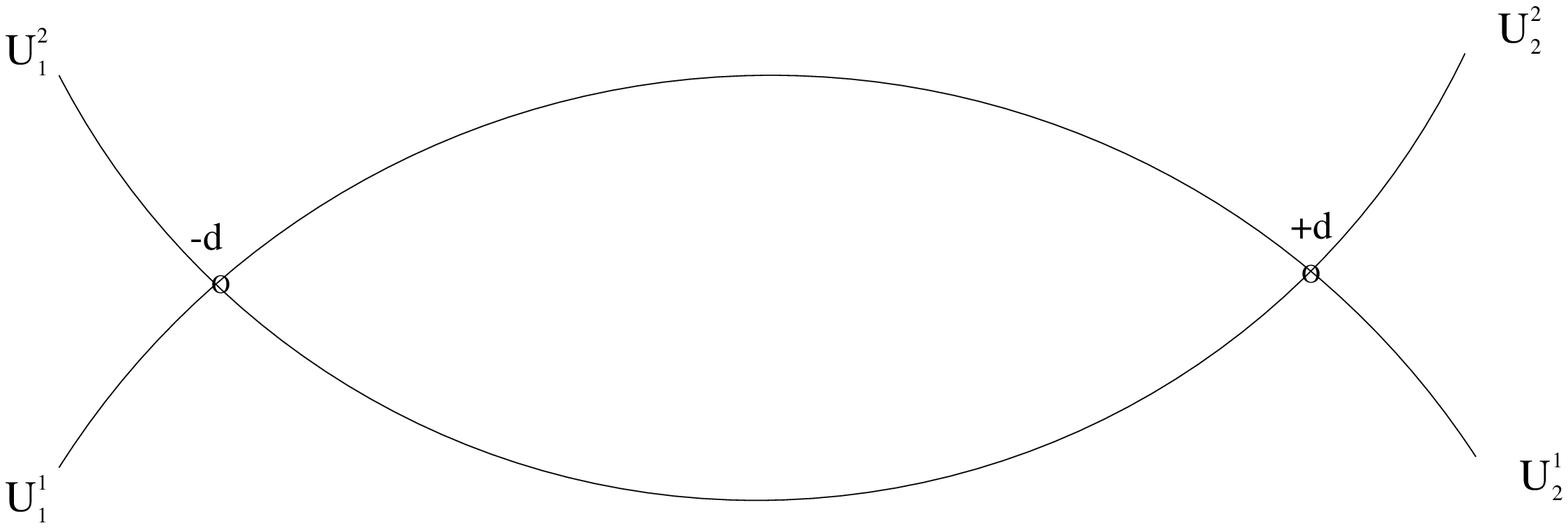}
\end{center}
\begin{itemize}
\item{ \bf Fig 3. } 
Two Luttinger liquids coupled together at two points ($x=-d$ and
$x=+d$) and connected to external reservoirs held at constant voltages
$U_1^1, U_1^2$ on the left and $U_2^1, U_2^2$ on the right.
\end{itemize}

Now, let us see how one  gets this result in a little more detail. We
start with spinless fermions, ( spin is an added complication, which
can be incorporated at a later stage), bosonise them and describe the
Luttinger liquid as 
\beq
H = {\hbar v_F\over 2 g}\Sum_{A=1}^2 \int dx [g(\partial_x\phi_A)^2 +
g^{-1}(\partial_x\theta^A)^2].
\eeq
The external voltage biases are incorporated as boundary conditions
on the boson fields at $-L$ and $+L$, where $L$ is the length of both
the wires. At the two coupling points, we
allow density-density couplings and single particle tunnelings. In
fact, for repulsive interactions, single particle tunnelings just
renormalise the density-density couplings, so we only  need to
introduce the interaction term given by
\beq
V_{\rm den} = \lambda_1 \rho^1(-d) \rho^2(-d) + \lambda_2 \rho^1(+d)
\rho^2(+d).  
\eeq
With standard normalisation, these operators have scaling dimension 
$2g$. Since for a bulk operator, relevance or irrelevance depends on
whether $g$ is less than or greater  than one, we see that for
$1/2<g<1$, 
these couplings are irrelevant and there is perfect 
transmission in both the wires - 
\beq
I^A = e^2 (U^A_1-U^A_2)/h \equiv  e^2 U^A/h.
\eeq
But for $1/2<g<1$, the operators are relevant. However, in this case, 
we find that the model can be mapped to decoupled
wires as shown in Fig.(4).
\epsfxsize=3.2 in
\epsfysize=2.0 in
\begin{center}
\epsffile{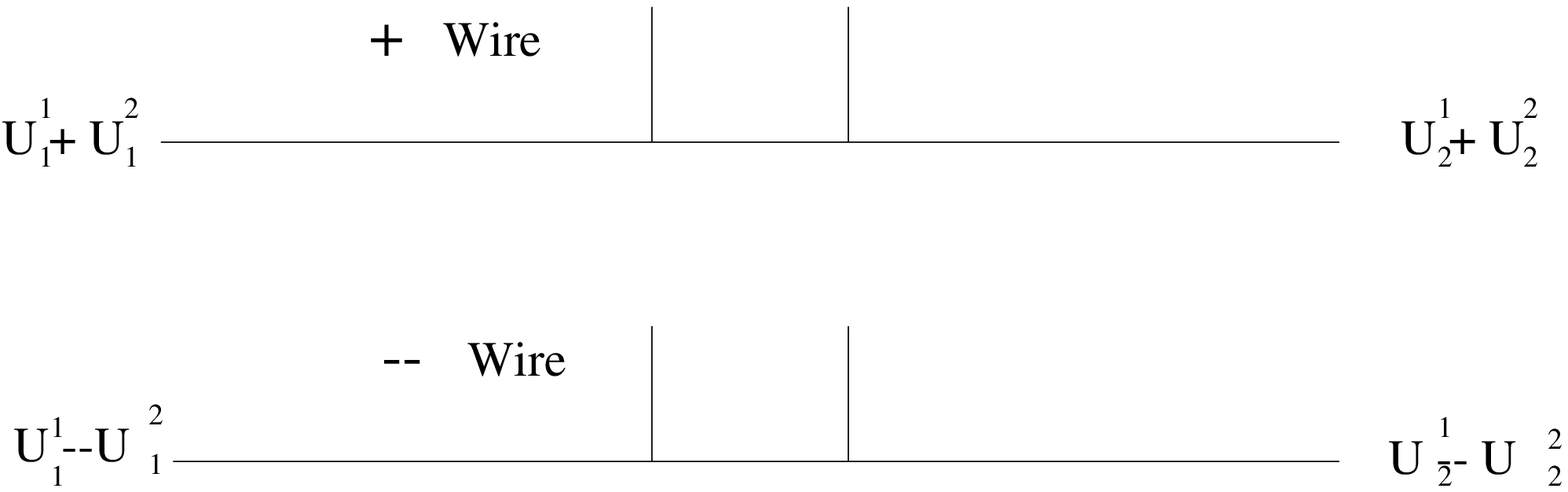}
\end{center}
\begin{itemize}
\item{ \bf Fig 4. } 
Two decoupled wires with barriers at ($x=-d$ and
$x=+d$). The values of the external voltages have changed.
\end{itemize}
The external biases have changed  so that the potential drops are now
$U^1+U^2$ in the $+$ wire and $U^1-U^2$ in the $-$ wire. The coupling
constant $g \longrightarrow {\tilde g} = 2g$ and hence the dimensions
of the barrier operators or the RG equations for the barriers have 
changed, but now the problem is easy to analyse because it maps
exactly into two copies of the Kane-Fisher problem. Hence, we can
directly take over their results. Their resonance condition for the two
wires ( with our changed parameters) is given by
\beq
4e g^2 \Delta \phi_G^{\pm} = \pi \hbar v_F/2d
\eeq
where $\Delta \phi_G^{\pm}$ is the spacing of the gate voltages.
In the Kane and Fisher analysis, they did not worry about the charging
of the barriers. This was later included\cite{CHARGING} 
by other workers in the
field. If we include those effects as well, we get
\beq
4e g^2 \Delta \phi_G^{\pm} = [{2d\over\pi \hbar v_F} +
{2C^{\pm}(2g)\over e^2}]^{-1}
\eeq
where $C^{\pm}$ is the capacitance of the barriers.
The inclusion  of the barrier capacitance decreases the spacing of
the gate voltages where we get resonances. One can understand this as
follows. The charging of the barriers increases the island's capacity
to hold charge. Since the spacing of the gate voltage is inversely
proportional to the capacitance of the island, this decreases the
spacing and we get more resonances within a given range of the gate
voltage. For strong barriers, the capacitances have no dependence on
the external biases, and $C^+ = C^-$.  So since the lengths of the two
wires between the barriers are also the same, ($d^+=d^-$), the + and -
wires satisfy the same condition for resonance and $\Delta \phi_G^+ =
\phi =
\Delta \phi_G^-$.

In terms of the original wires 1 and 2, resonance implies 
the condition that the current in each of the wires before the
crossing is equal to that after the crossing - $i.e.$, $I_1=I'_1$ and
$I_2=I'_2$. We can consider two cases. 
Let us first consider the case when one of the wires, say 
wire 2 is unbiased and
$I_2=I'_2=0$. We want to look for when there is resonant tunneling
through wire 1.  We see that when the
resonance condition is satified in both the + and $-$ wires, we get $I^+ =
I^-$, which in turn gives us $I_1=I'_1$,  since there
is no current in wire 2. We can also consider the case when one of the
decoupled wires is unbiased, for instance, the $-$ wire is unbiased. In
that case, we have the same current flowing through both the wires -
$i.e.$, $I_1=I'_1=I_2=I'_2$ when $\Delta\phi_G = \Delta \phi_G^1 
+\Delta \phi_G^2 = 2\phi$ is a constant. In this case, resonant
tunneling takes place through both wires. Note that even in case 1
where tunneling only occurs in one wire, the situation is still
different from that of two originally decoupled wires, because the resonance
condition has changed from that of a single wire.

In general, without resonant transmission, $I_1\ne I'_1$ and $I_2\ne
I'_2$. So one needs four current probes to measure the current
characteristics, in terms of a four by four matrix.

\vskip 0.3cm

\noindent {\bf Transport in an exactly solvable toy model of Luttinger
liquid}

\vskip 0.3cm

Recently, a toy model of a Luttinger liquid was proposed by Schulz and
Shastry\cite{SS}. 
It is a model with two species of fermions with a pseudospin
index $\sigma =\pm$, and a Hamiltonian given by
\beq
H=\sum_{i\sigma} (p_{\sigma i} + \sigma A_{\sigma} (x_{-\sigma i}))^2
\eeq
where $A_{\sigma}$ is a gauge potential which for a positive
pseudospin particle depends on all the negative pseudospin particles
and vice-versa - $i.e.$, 
\beq
A_{\sigma}(x) = \sum_j V(x-x_{-\sigma j}).
\eeq
The mainpoint is that since the potential depends on the total number
of particles of the opposite pseudospin, every time a + particle is
added, the energy levels of all the $-$ particles change and vice versa.
This model is easy to solve because one can make a gauge
transformation to remove the gauge field so that
$$H \longrightarrow \sum_{\sigma i} p_{\sigma i}^2$$
at the expense of changing the boundary conditions on the
wave-functions. So if we take the particles to be on a ring, instead
of quantising $k_i = 2\pi n_i/L$, the changed bundary conditions lead
to the changed quantisation condition given by  
\beq
k_{\pm i} = {2\pi\over L} (n_{\pm i} \pm {N_{\mp} \delta\over 2 \pi})
\eeq
where $N_{\mp}$ is the total number of particles of the $\mp$ in the
wire. Clearly the Hilbert space
of states for $N_{\mp}\delta/2\pi$ = fractional 
is different from that of a non-interacting model with  
$N_{\mp}\delta/2\pi$ = integer. Hence, even though, it `looks' like a
non-interacting theory, as far as the Hamiltonian is concerned, the
changed boundary conditions incorporate the interactions of the
model that existed before gauge transformation.
In the original paper, they computed the
correlation functions in this model and showed that they could get
fractional exponents, which is the hallmark of Luttinger liquids.
With the motivation of studying coupled chains of Luttinger liquids,
we tried to generalise this model. However, the natural generalisation
led to a model which was more like a multi-band single chain model,
which we analysed\cite{RKG} and obtained correlation functions.

Currently, we are studying transport in this model\cite{PEEKAY}. 
Since interactions
in this model can only be introduced through a change in boundary
conditions and consequently quantisation conditions, we study the
model on a ring. Hence, the driving force is a flux through the ring
rather than external voltage sources as for an open wire. Like for
free fermions, we expect to get persistent currents. On explicitly
introducing barriers (potentials), we expect to see results similar
to those in the  Kane-Fisher model. Repulsive interactions will 
cut the wire and
attractive interactions will heal the wire. However, here, since mesoscopic
length scales are involved, the cutting and healing may not be perfect. 

In our opinion, the importance of this model lies in the fact that even
in the original fermion language, the model is almost free, with
interactions only being introduced through quantisation
conditions. Hence, it should be possible to get results for the model
and consequently for a Luttinger liquid, without going through
bosonisation. We are hence, trying to see whether we can reproduce the
Kane-Fisher results on the ring, without going through bosonisation in
this model.

\vskip 0.3cm

\centerline{\bf Conclusions}

\vskip 0.3cm

In conclusion, I would like to emphasize that the field of transport
in Luttinger liquids is a highly relevant field at the moment,
because a lot of experiments are likely to be performed in the near future on
wires operating in the single channel limit, on carbon nanotubes,
etc. Hence, the Landauer-Buttiker formalism for mesoscopic wires needs
to be redone for these strongly interacting electrons or for the
Luttinger bosons. 

There are other interesting phenomena in this general area, 
which we have not touched upon
in this talk. For instance, inclusion of spin will lead to the
formation of Kondo resonances. Inclusion of AC voltages can lead to
novel phenomena. New materials are constantly being made, which could
have new physics. Examples are the amchair carbon nanotubes, and the
chiral nanotubes. 

Hence, both at the theoretical level and at the experimental level, we
expect the field to expand considerably in the near future.

\section*{Acknowledgments}
I would like to thank my collaborators P. Durganandini and
P.K. Mohanty for many useful conversations.

\end{document}